\documentclass[%
 preprint, 
 amsmath,amssymb,
 aps, physrev,
]{revtex4-2}

\usepackage{xcolor}
\usepackage{array}
\usepackage{graphicx}
\usepackage{dcolumn}
\usepackage{bm}

\begin{document}

\preprint{APS/123-QED}

\title{\textbf{An Analysis of Stable Mode Contributions to Rayleigh-B\'{e}nard Convection} 
}

\author{Zachary R.~Williams}
 \email{Contact author: williamsz@hope.edu}
\author{Ava G.~Lowe}%
\affiliation{ Department of Physics, Hope College, Holland, MI, USA 
}%

\date{\today}

\begin{abstract}
In this work, we investigate the presence and impact of stable eigenmodes in Rayleigh-B\'{e}nard convection that arises due to a temperature gradient within a fluid system. The nonlinear evolution of the canonical convection system is cast in terms of eigenmode amplitudes. The linear modes that play a significant role in the dynamics across a range of Rayleigh numbers are identified. We find that while unstable eigenmodes are a significant contributor, a small number of linearly stable modes grow to large amplitude via nonlinear interactions and are essential in modeling system dynamics. Importantly, the stable eigenmodes are seen to contain the majority of the boundary layer structure that forms in the nonlinear state. We also demonstrate that the scaling of convective heat transport (as quantified by the Nusselt Number) with Rayleigh number can be effectively quantified with only a small fraction of the total number of eigenmodes in the system. 
\end{abstract}

\maketitle


\section{Introduction}
The problem of {Rayleigh-B\'{e}nard Convection\cite{RBC} (or natural convection) is ubiquitous to fluid systems, providing a description of dynamics ranging from atmospheric fluid dynamics\cite{Emanuel94}, terrestrial flows \cite{Calkins12,Schubert01}, astrophysical systems\cite{Spiegel71,Getling98}, and even fusion plasmas\cite{Wilczynski19}. While this process has been thoroughly studied for over one hundred years, there still remains much to be learned about broader fluid dynamics ideas through more investigation. While in principle all the information to be gleaned about RBC systems lie within the solutions to the governing nonlinear equations, solutions at extreme parameters remain analytically unsolvable and computationally intractable. Rather than examining thermal convection through further extensive nonlinear simulations, this work proposes to analyze well understood convective processes through the lens of eigenmode decomposition to deepen understanding.}

{There exists a growing collection of works that demonstrate the physics insights offered through eigenmode decomposition analysis, cf.\cite{Baver02,Terry06,Hatch11,Mawkana11,Fraser17,Stolnicki24,Fraser18} The study of linear eigenmodes is commonly used to characterize the instability present in a given equilibrium. However, these works highlight the fact that nonlinear behavior in a variety of systems depend not only on unstable eigenmode dynamics but also on linearly stable (or damped) modes, which become significant through nonlinear interactions. These nonlinear interactions produce not only Kolmogorov-like cascades of energy from large to small spatial scales, but local-in-wavenumber interactions can produce stable modes as a significant source/sink of energy even without considering a cascade.}

{It is also noteworthy that in many of these works, eigenmode decomposition analysis serves to significantly reduce the degrees of freedom needed to adequately describe a system, even in a fully developed nonlinear state. In certain contexts, the essential dynamics of a fluid system can be captured using a small fraction of the total number of eigenmodes present (as few as two or three in particularly exceptional cases). For systems in which the degrees of freedom are significantly constrained and require a very small number of eigenmodes, new physics insight can be achieved without the need for fully-evolved, computationally expensive nonlinear simulations. In some instances the fact of local unstable-stable eigenmode interactions allows for reduced model developments and new physics otherwise untenable in a fully nonlinear description\cite{Hegna18,Fraser18}}

{We examine the canonical problem of Rayleigh-B\'{e}nard convection in anticipation of new insights that it may reveal for this system. By using the linear eigenmodes as a basis for the vector space of nonlinear solutions, we show that significant nonlinear physics is intimately connected to the behavior of specifically the linearly damped eigenmodes. The boundary layer dynamics, a characteristic feature of thermal convection, is attributed primarily to the structure of stable eigenmodes only. Additionally, how the Nusselt number scaling with Rayleigh number, a classic question in the study of convection, is accurately reproduced using a substantially-constrained, reduced-ordered system consistent of only 10\% of the overall eigenmodes. This results are observed to hold robustly across several orders of magnitude in Rayleigh number.} 

{This paper is organized as follows. In Section \ref{methods}, we discuss the mathematical used to describe thermal convection, the computational setup for our studies, and the eigenmode decomposition analysis employed in this and other words. In Section \ref{eigens} we discuss the characterization of the linear eigenmodes and their time-evolved amplitudes, highlighting the eventual dominance of stable modes that is ubiquitous across a range of Rayleigh numbers. Section \ref{NLs} explore aspects of nonlinearly-evolved convection that can be attributed specifically to stable mode dynamics as well as a reduced-order scaling of Nusselt number with Rayleigh number. We then summarize and discuss the work, including future directions, in Section \ref{Summary}}.

\section{\label{methods}Methods}

{Rayleigh-B\'{e}nard convection is studied in this work using the Dedalus code\cite{Burns20} (following the approach similar to works such as Refs.~\cite{Anders18,OConnor21}) The specific model employed is the Boussinesq approximation of the Navier-Stokes equation\cite{Chandrasekhar81} given below:}

\begin{align}
    \vec{\nabla}\cdot\vec{u} &= 0 \label{NS1} \ \ ,\\
    \frac{\partial \vec{u}}{\partial t} + \vec{u}\cdot\vec{\nabla}\vec{u} &= -\vec{\nabla}\varpi + T\hat{z} + \mathcal{R}\nabla^2\vec{u} \label{NS2}\ \ ,\\
    \frac{\partial T}{\partial t} + \vec{u}\cdot\vec{\nabla}T &= \mathcal{P}\nabla^2\vec{T} \label{NS3} \ \ . \\
\end{align}

{Characteristic parameters of the system include the kinematic viscosity $\nu$, the thermal diffusivity $\kappa$, and the thermal expansion coefficient $\alpha$. Standard gravitational acceleration is assumed $\vec{g} = -g\hat{z} = -\vec{\nabla}\phi$ with a corresponding gravitational potential $\phi$. These equations are normalized in the following way: lengths are normalized with respect to the layer height $(L_z)$, velocities are normalized with respect to freefall velocity $(v_{ff} = \sqrt{\alpha g L_z^2\partial_zT_0})$, time is normalized to freefall time $(t_{ff} = L_z/v_{ff})$, and temperature is normalized to the equilibrium temperature different across the layer $(\Delta T_0 = L_z\partial_zT_0)$. The fluid velocity is assumed 2D, $\vec{u} = u\hat{x} + w\hat{z}$. Pressure and gravitational effects are combined in a reduced kinetmatic pressure ($\varpi \equiv P/\rho_0 + \phi$), which is normalized by $v_{ff}^2$. Dimensionless quantities are defined in terms of the Rayleigh and Prandtl numbers (Ra and Pr, respectively): $\mathcal{R} = \sqrt{\mathrm{Pr}/\mathrm{Ra}}$, $\mathcal{P} = 1/\sqrt{\mathrm{Pr} \mathrm{Ra}}$, $\mathrm{Ra} = (L_z v_{ff})^2/\nu\kappa$, $\mathrm{Pr} = \nu/\kappa$. For the entirety of this work, Prandtl = 1 is assumed.}

{All simulations discussed in this work are two-dimensional ($x$ and $z$, $y$ is the ignorable coordinate). While there are importance differences in Rayleigh-B\'{e}nard Convection that arise when considering 3D effects\cite{VanderPoel13}, examination of such effects are beyond the scope of this paper. $x$ and $z$ directions refer to the homogeneous and inhomogeneous directions, respectively. The domain aspect ratio $L_x/L_z = 4$ is held fixed throughout this work and the domain ranges from $(0,L_x)$ and $(-L_z/2, L_z/2)$. No-slip, fixed-temperature boundary conditions are assumed in the $z$ direction, such that $u = w = T = 0$ at $z = -L_z/2$ and $z = L_z/2$. Periodic boundary conditions are used for the $x$ direction. Simulations are conducted at a variety of Rayleigh numbers, ranging from $10^5$ to $10^8$. $x$ and $z$ resolutions, $N_x$ and $N_z$, for the different Rayleigh numbers are given in Table \ref{param_table}. Unless otherwise mentioned, all simulations discussed in this paper assume a diffusive equilibrium of the form of a linearly decreasing temperature profile $T_0(z) = \frac{L_z}{2} - z$. For direct simulations of convective dynamics, an initial perturbation is introduced as random noise and allowed to evolve into a steady state (roughly 200 freefall units of time).}

\begin{table}[b]
\caption{\label{param_table}%
{Resolutions used for each value of Rayleigh number discussed in this paper.}
}
\begin{tabular}{||m{1.5cm} || m{1.5cm} | m{1.5cm} | m{1.5cm} | m{1.5cm} | m{1.5cm} | m{1.5cm} ||} 
\hline
 $Ra$ & $10^5$ & $5\times10^5$ & $10^6$ & $5\times10^6$ & $10^7$ & $10^8$\\ 
 \hline
 $N_x$ & 256 & 256 & 256 & 512 & 512 & 1024 \\
 \hline
 $N_z$ & 64 & 64 & 128 & 256 & 256 & 512 \\
 \hline
\end{tabular}
\end{table}

{A brief review of the eigenmode decomposition analysis technique is given in following text (see works such as Refs.~\cite{Baver02,Terry06} for a more complete discussion). The nonlinear evolution of a perturbation $f(x,y,z,t)$ in some fluid quantity of interest (e.g. flow velocity or temperature) can be described with the following schematic equation:} 

\begin{equation}\label{cartoon_eq}
{\frac{\partial f}{\partial t} = \mathcal{L}[f] + \mathcal{N}[f] \ ,}
\end{equation}

\noindent {where $\mathcal{L}$ and $\mathcal{N}$ correspond to the linear and nonlinear portions of the governing equations. Equation \ref{cartoon_eq} could represent the Navier-Stokes equations for fluid dynamics or the Boltzmann Equation in a kinetic description, as examples. By definition, the linear term is first order in $f$ and the nonlinear term is higher order. Thus, for small values of $f$ (corresponding to perturbations from an equilibrium state) the nonlinear term can be neglected and the resultant linear equation can be solved. The solutions for the linearized equation are referred to as the eigenmodes $f_j$ of the system. Eigenmodes are typically characterized as exponentially growing (unstable), exponentially decaying (stable), or constant (marginal) in time. These eigenmodes form a complete basis meaning that any general solution $f$ to the full nonlinear equation can be expressed as a linear combination of the eigenmodes:} 

\begin{equation}\label{eigenmode_expansion}
{f(\mathbf{r},t) = \sum_j \hat{\beta}_j(t)f_j(\mathbf{r})  =  \sum_{k_x}\sum_j \beta_j(k_x,t)\hat{f}_j(z)e^{ik_xx} \ .}
\end{equation}

\noindent {The first equality represents the eigenmode expansion, the second equality represents a standard Fourier expansion in the periodic direction ($x$ in this work) that will be utilized in subsequent discussion. Note that the validity of the eigenmode expansion is a consequence of the specific model being examined and requires that the full set of eigenmodes $f_j$ form a complete basis. This property has been confirmed numerically for all systems to be investigated in this work. In Equation \ref{eigenmode_expansion} the full nonlinear solution is expressed as a function of both space and time. As the eigenmodes $f_j$ form a complete (though not necessarily orthogonal) basis, Equation \ref{eigenmode_expansion} is an exact expression given appropriately-determined coefficients $\beta_j$. Through the use of left-eigenvectors $f_{\mathrm{L},j}$ (which are always orthogonal to right eigenvectors) and a suitably-defined inner product (represented by brackets in the equation below), these coefficients can be calculated:} 

\begin{equation}\label{beta_calc}
{\beta_j = \frac{\langle f^*_{\mathrm{L},j} | f \rangle}{\langle f^*_{\mathrm{L},j} | f_j\rangle} \ ,}
\end{equation}

\noindent {where the asterisk denotes a complex conjugation (similar to Dirac notation in quantum mechanics where the complex conjugate is implied). For computational studies, the total number of modes and thus number of terms in the summation described by Equation \ref{eigenmode_expansion} depends on the numerical resolution and in general is a large number (on the order of thousands). As such, working with the full summation over all of the eigenmodes does not offer any special or practical utility beyond directly finding the nonlinear solution. However, many studies using this eigenmode decomposition approach have found that particular eigenmodes stand out as having especially large coefficients and thus provide dominating contributions to the system dynamics. It is especially noteworthy that for most of these studies, thorough investigation finds significant contributions from modes that are linearly stable and would often be ignored in quasilinear models or other theoretical descriptions. Similar to previous efforts (cf. Refs.~\cite{Terry06,Mawkana11,Hatch11,Fraser17}), the significance of each eigenmode in a nonlinearly-evolved system is quantified using the $\beta_j$ coefficients defined above, hereafter referred to as the \textit{eigenmode amplitude}.}

\section{Eigenmode Characterization\label{eigens}}

{Before the contribution of individual eigenmodes to the nonlinear state can be discussed, the eigenmodes themselves must be determined. As described in the previous section, these eigenmodes are determined by solving the linearized versions of Equations \ref{NS1}-\ref{NS3}, given below:} 

\begin{align}
    \vec{\nabla}\cdot\vec{u_1} = 0 \label{linNS1} \ \ ,\\
    \frac{\partial \vec{u_1}}{\partial t} + \vec{\nabla}\varpi_1 - T_1\hat{z}  - \mathcal{R}\nabla^2\vec{u_1} = 0    \label{linNS2}\ \ ,\\
    \frac{\partial T_1}{\partial t} + w_1\frac{\partial T_0}{\partial z} - \mathcal{P}\nabla^2\vec{T} = 0   \label{linNS3} \ \ . \\
\end{align}

\noindent {The ``1'' and ``0'' subscripts refer to fluctuating and equilibrium quantities, such that $f(x,z,t) = f_0(z) + f_1(x,z,t)$ where $f$ represents any dynamic variable. In linearizing the equations it is assumed that $f_1 \ll f_0$, and that there is an equilibrium temperature profile as discussed previously and no equilibrium fluid flow. Fluctuating quantities take the assumed form of $f_1(x,z,t) = f_1(z)\mathrm{exp}\left[i(k_x x - \omega t)\right]$, where $k_x$ represents Fourier wavenumber in the periodic direction and $\omega = \omega_r + i\gamma$ consists of a real frequency $\omega_r$ and growth rate $\gamma$. The different Fourier wavenumbers $k_x$ are decoupled from one another linearly. Given a choice of $k_x$, Equations \ref{linNS1} - \ref{linNS3} form an eigenvalue problem to determine $\omega_j$ and $f_j(z)$ for all eigenmodes $j$.}

\begin{figure}
    \centering
    \includegraphics[width=0.75\linewidth]{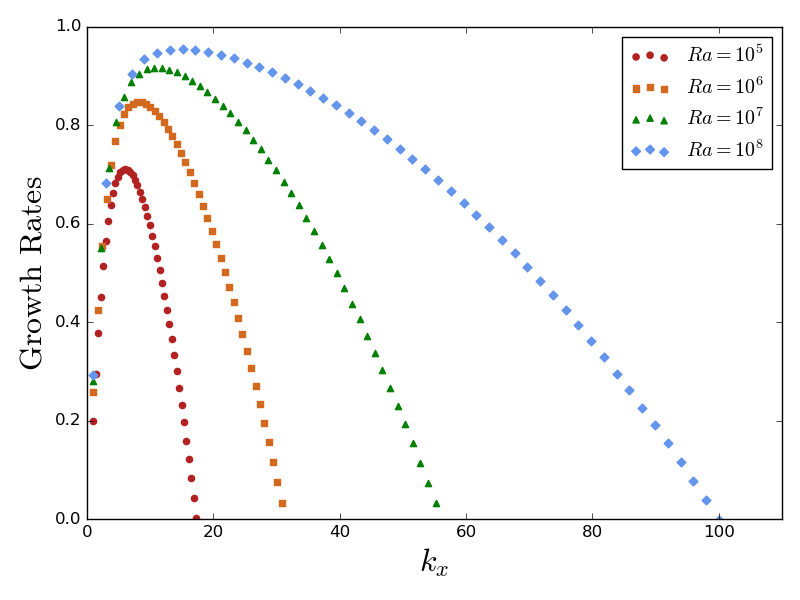}
    \caption{Maximum unstable growth rate vs wavenumber $k_x$ over several orders of magnitude in Rayleigh number. At a given $k_x$ there are typically multiple unstable eigenmodes and the number of unstable eigenmodes at a given $k_x$ increases with Rayleigh number, as does the wavenumber range for instability.}
    \label{fig:growthrates}
\end{figure}

{Figure \ref{fig:growthrates} shows eigenmode solutions with the largest growth rates as a function of wavenumber $k_x$ across a range of Rayleigh numbers. Across all parameters, the spectrum of eigenmodes all have exactly zero real frequency. They are thus characterized as either uniformly growing or uniformly decaying. As Rayleigh number increases, so do both the range of unstable Fourier wavenumbers as well as the number of unstable eigenmodes, although the number of stable modes vastly exceeds that of unstable modes at each wavenumber.}

\begin{figure}
    \centering
    \includegraphics[width=0.75\linewidth]{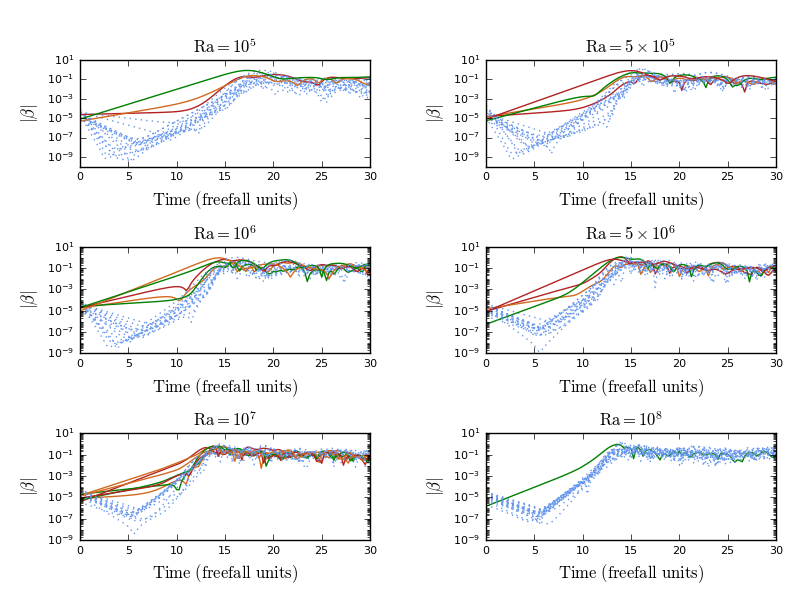}
    \caption{Eigenmode amplitude evolution over time for a range of Rayleigh number at the $k_x$ for which the linear growth rates peak. Each simulation was run to a total of 200 freefall time units, the majority of the evolution consisting of a quasi-stationary state. The plots are truncated to aid in visualization of early (linear) stage of evolution. Only the top twenty eigenmodes are shown for reach value of Ra, as ranked by their amplitudes time-averaged over the saturated state. Solid lines represent linearly unstable modes, dashed lines represent linearly stable modes. In each case, local nonlinear interactions enhance linearly stable eigenmodes to amplitudes that are comparable to the unstable modes. Importantly, for many values of Ra there are linearly unstable modes that do not fall in the top twenty based on eigenmode amplitude. Ra = $10^8$ represents an extreme case of this, containing the largest number of unstable modes yet the fewest that rank among the largest amplitude.}
    \label{fig:betas}
\end{figure}

{Equipped with eigenmode solutions, contributions of each eigenmode to a given nonlinear time evolution can be assessed by calculating amplitudes as given in Eq.~\ref{beta_calc}. In all of the following discussion, unless otherwise stated, these analyses are performed locally in wavenumber (at a specific $k_x$, representing a single term in the Fourier summation of Eq.~\ref{eigenmode_expansion}). The $k_x$ value is selected such that it falls most closely to the peak in linear growth rate for a given Rayleigh number. The results of these calculations are shown in Fig.~\ref{fig:betas}. The same general features of the eigenmode amplitude time evolution holds across all Rayleigh number. For early times, the linear approximation is valid and the amplitudes behave as expected with unstable modes increasing in time and stable modes decreasing. However, once the unstable modes grow to significant amplitude, nonlinear coupling transfers energy from unstable modes to stable modes (as original described in Ref.\cite{Terry06}) and many of the stable mode amplitudes begin to grow. Importantly, this transfer of energy (mediated via a three-wave-interaction involving the mean flow) occurs at the same wavenumber, rather than involving a cascade of energy to different spatial scales. This demonstrates that stable modes can serve as a local sink of energy to saturate unstable mode growth in contrast to just relying on a forward cascade to small-scale dissipation effects. As the simulation progresses past the linear phase and into nonlinear, many of the stable eigenmodes reach an amplitude comparable to or even exceeding that of the unstable mode. The eigenmodes can be ranked by their respective amplitudes as averaged over the quasi-stationary state, revealing that stable modes are among the most significant across all Rayleigh numbers, and in fact for certain cases stable modes completely dominate the picture (refer specifically to Ra = $10^8$).} 

{It should be noted the eigenmode amplitudes can exhibit sensitivity to the initial conditions. As such, it can be argued that these results are specific to the initial condition chosen here and not a more general consequence. To test this, the analysis was repeated initializing the nonlinear simulations with the exact structure of the most unstable eigenmode, using Ra = $10^5$ as a representative case. While the initial linear evolution differed from the case shown here (as expected), the resultant nonlinear quasi-stationary state was qualitatively identical, with stable modes growing to significant amplitude despite the preferential designation for unstable modes in the initial condition. We thus conclude that the tendency for nonlinear interactions to enhance stable mode activity to comparable amplitudes as the unstable modes is a general feature of this system, similar to other eigenmode decomposition analyses, cf.~\cite{Fraser21,Stolnicki24}.}

\section{Identifying Stable Mode Contributions \label{NLs}}

{This section examine specific aspects of fully evolved convective simulations, specifically focusing on the role of specific eigenmodes or groups of modes. One characteristic feature of thermal convection subject to these specific boundary conditions is the formation of a boundary layer in which rapid variation of the fluid velocity and temperature occurs in contrast to the relatively flat bulk profiles. Thermal and kinetic boundary layers (defined via sharp variations in temperature and flow, respectively) play a substantial role in determining nonlinear properties of RBC. For Pr$\sim$ 1, the size of the thermal and kinetic boundary layers are of the same order and so in this work we will refer specifically to the thermal boundary layer as representative of general boundary layer features.}

{Eigenmode amplitudes allow for a separation of contributions to the boundary layer structure between unstable and stable modes. At a given Fourier wavenumber $k_x$, following Eq.~\ref{eigenmode_expansion}, the $z$ variation in temperature can be expressed as}

\begin{equation}\label{split_eigensum}
{T(z,t) = \sum_{j, \gamma_j>0} \beta_j(t)\hat{T}_j(z)e^{ik_xx} + \sum_{j,\gamma_j<0} \beta_j(t)\hat{T}_j(z)e^{ik_xx} \ \ \ ,}
\end{equation}

\begin{figure}
    \centering
    \includegraphics[width=0.75\linewidth]{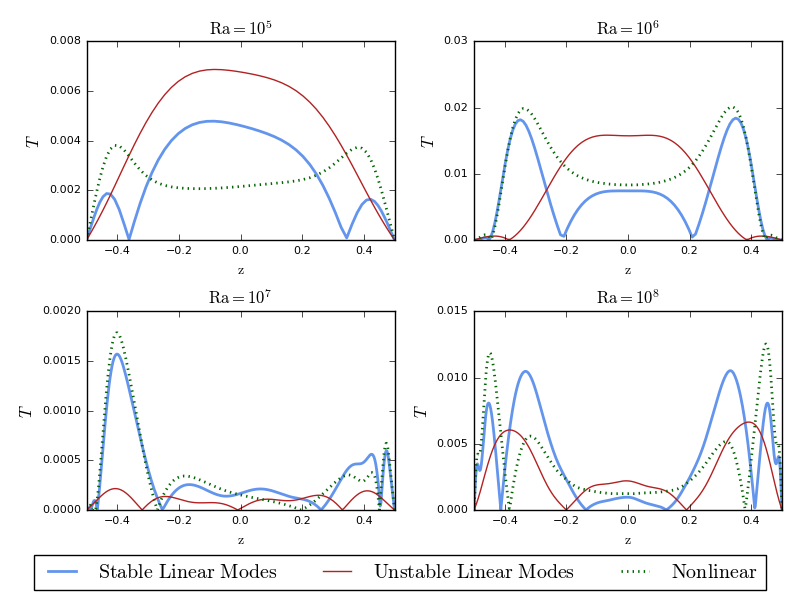}
    \caption{Time-averaged temperature fluctuation profiles at the $k_x$ wavenumber corresponding to peak linear growth rates. Dashed green lines represent profiles as determined from the nonlinear data, solid blue lines represent profiles determined from only the stable mode contributions, and solid red lines represent contributions from only the unstable modes (see Eq.\ref{split_eigensum}). The boundary layer is evident in the nonlinear profile, marked by the significant variations near the edges of the $z$ domain. For all Rayleigh numbers considered, the variation in structure near the domain edges is captured primarily in the stable modes, rather than unstable. This is consistent with the observed large-amplitude stable mode activity and further highlights stable mode importance in nonlinear system dynamics.}
    \label{fig:Blayers}
\end{figure}

\noindent {where the first term in the sum represents contributions from only the linearly unstable modes and the second term contributions from the linearly stable modes. The $z$ dependent temperature profiles are determined from a time-average over the quasi-stationary state in the evolution as well as from the summations in Eq.~\ref{split_eigensum} where the amplitudes $\beta_j$ are averaged over the quasi-stationary state as well. The results for a range of Rayleigh numbers are shown in Fig.~\ref{fig:Blayers}. This separation into unstable/stable contributions reveals that the steepest changes in the time-average $T(z)$ in the boundary layer are contained in the linearly stable modes, not the unstable modes. A model constructed only on the initial equilibrium's unstable eigenmodes would thus omit a significant portion of the boundary layer structure and its effects on convection.}

\begin{figure}
    \centering
    \includegraphics[width=0.75\linewidth]{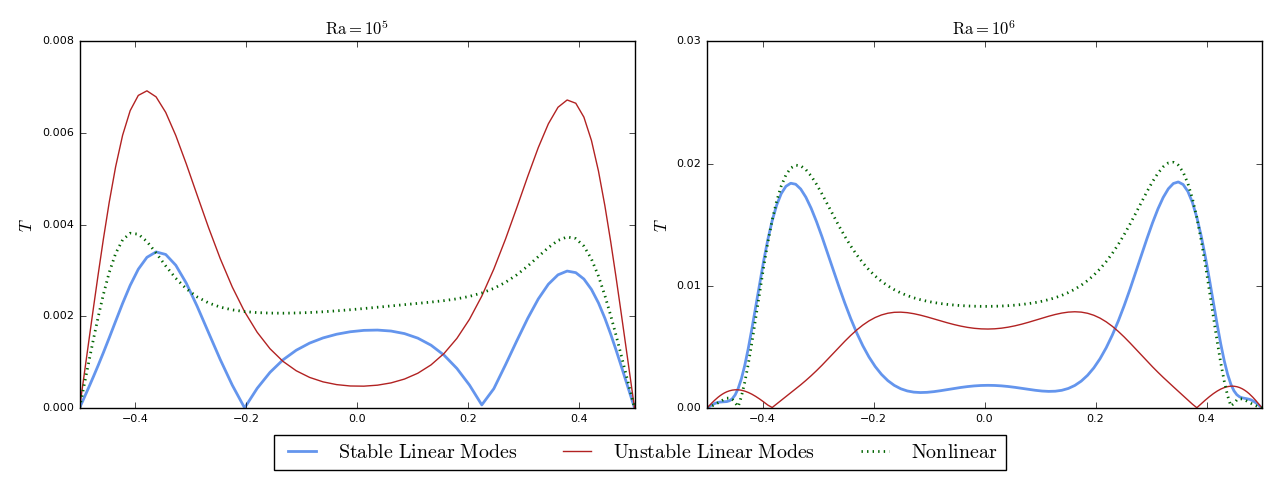}
        \caption{Time-averaged temperature fluctuation profiles as discussed in Fig.~\ref{fig:Blayers}, differing in eigenmode basis. For these results, the eigenmodes were calculated from an equilibrium based on the nonlinearly-evolved temperature profiles. These unstable modes do contribute more significantly to boundary layer features. However, it is evident that stable modes are also essential in effectively capturing the boundary layer.}
    \label{fig:Blayers2}
\end{figure}

{Another reasonable question to consider within this discussion is the matter of which set of eigenmodes one chooses to use in describing a nonlinear state via Eq.~\ref{eigenmode_expansion}. All the results here use as a basis the eigenmodes of the initial diffusive equilibrium (linear variation in $T_0(z)$). One can argue that a more appropriate set of eigenmodes would be those that arise from the nonlinearly-evolved profiles. To investigate this claim, linear analyses at Ra = $10^5$ and $10^6$ were performed using Eqs.~\ref{linNS1} - \ref{linNS3} and the time-averaged $T(z)$ from the nonlinear quasi-stationary state as the input $T_0(z)$ profile. This produced a new set of eigenmodes as a basis to repeat the amplitude calculations and subsequent analysis. As evident in Fig.~\ref{fig:Blayers2}, these unstable modes do contribute more significant than those based on the diffusive equilibrium. However, it is clear that the stable modes still contribute substantially to boundary layer structure. We thus maintain that the qualitative conclusions drawn from the diffusive equilibrium and that the original eigenmodes are a physically meaningful basis from which to build the model.}

\begin{figure}
    \centering
    \includegraphics[width=0.75\linewidth]{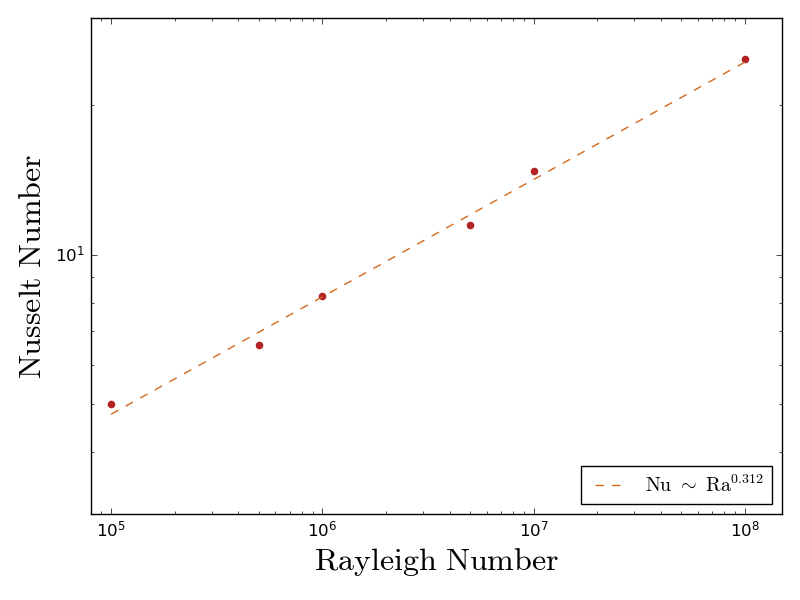}
    \caption{Nusselt number as a function of Rayleigh number, as calculated from a truncated eigenmode expansion retaining only the top 10\% of eigenmodes as ranked by time-averaged amplitudes. The dashed line represents a fit used to determine the scaling relationship, revealing a scaling of roughly Nu $\sim $ Ra $^{3.12}$. This is consistent with existing models, thus demonstrating the effectiveness of the truncated eigenmode expansion in describing the physical properties of the system utilizing far fewer degrees of freedom.}
    \label{fig:NuRa}
\end{figure}

{Another fundamental question in thermal convection is the matter of Nusselt number scaling with Rayleigh number. The Nusselt number quantifies the amount of convective vs. diffusive heat transport in a system, defined as  }

\begin{equation}
{\mathrm{Nu} = \frac{\mathrm{convection} + \mathrm{diffusion}}{\mathrm{diffusion}} = \frac{\langle wT - \mathcal{P}\partial_zT\rangle}{\langle-\mathcal{P}\partial_zT\rangle} = 1 + \sqrt{\mathrm{Ra}}\langle wT \rangle} \ \ .
\end{equation}
{The last equality applies the fact that Pr = 1 for this work, and the brackets denote volume average: $\langle f\rangle = \int_0^{L_x}\int_0^{L_z} f dz dx$. Direct calculation of Nusselt number from fully evolved nonlinear simulations is straightforward. However, this discussion prompts an investigation of the aforementioned claim of eigenmode analysis's ability to reduce the necessary degrees of freedom needed to a nonlinear system effectively. The Nusselt number is calculated by determining $w$ and $T$ from eigenmodes as described in Eq.~\ref{eigenmode_expansion}. However, rather than a summation over all of the eigenmodes (which would exactly reproduce the Nusselt number from a direct nonlinear simulation), a summation was performed over only the top 10\% of eigenmodes as ranked by their time-average $\beta_j$ values. This calculation yields Nusselt number as a function of time, which is then time-averaged over the quasi-stationary state. Results of this calculation as a function of Rayleigh number are shown in Fig.~\ref{fig:NuRa}. To compare the effectiveness of utilizing a truncated eigenmode sum to calculate Nusselt, a linear fit is applied to extract Nusselt scaling with Rayleigh number. The result of fitting Nu $\sim$ Ra$^\alpha$ produces $\alpha = .312$, which falls quite close to the commonly-accepted value of approximately 1/3 for this range of Rayleigh numbers\cite{Grossman00,OConnor21,Herring63,Kraichnan62} This highlights the fact that the system can be described to a reasonably high degree of accuracy using only a small subset of the total eigenmodes/degrees of freedom. Consistent with the prior discussion, a significant number of these top 10\% of eigenmodes used are linearly stable.}

\section{\label{Summary}Summary \lowercase{and} Discussion}

In this paper, we describe the methodology of eigenmode decomposition analysis and apply it to the problem of Rayleigh-B\'{e}nard Convection. A particular focus of this analysis is on the role that linearly stable eigenmodes play in the fully developed nonlinear state. This is achieved through the calculation of eigenmode amplitudes which reveal that stable modes, while decaying during the linear phase of the evolution, grow via nonlinear interactions to large amplitudes, often as large as or exceeding that of the linearly unstable modes.

These observations are not merely an academic curiosity but point to deeper insights as to the physical processes underpinning thermal convection. All of the analyses discussed in this paper occur at a single Fourier wavenumber, suggesting that this observed nonlinear transfer of energy to stable modes occurs locally in wavenumber space rather than via a cascade of energy across disparate spatial scales. This provides an additional avenue of saturation for unstable eigenmodes by transferring energy to stable modes at the same wavenumber via the mean ($k_x = 0$) profiles. 

Thermal boundary layers are also examined through the eigenmode decomposition lens. Stable mode $z$ profiles contain a significant amount of the temperature variations near the edges of the domain across a range of Rayleigh numbers, cementing the essential role they play in nonlinear convection models. These results do naturally exhibit sensitivity to choice in eigenmode basis. In utilizing an eigenmode basis as determined from the nonlinearly-evolved profiles, unstable mode structure contains more of the boundary layer variation. However, stable modes still contribute non-trivially and so we conclude the nonlinear enhancement of stable modes to be essential to boundary layer dynamics in either basis. 

Finally, we examine how important features of thermal convection, here specifically the Nusselt number, can be accurately captured using a significantly truncated description. While the total number of eigenmodes present in the numerical solutions (ultimately a consequence of system resolutions) is in the thousands, we find that retaining only 10\% of the total number of eigenmodes (most of which are linearly stable) produces a Nusselt/Rayleigh number scaling that matches current theories. This demonstration of how few basis elements are needed to effectively describe nonlinear properties allows for future reduced model developments (cf.~\cite{Hegna18} for an example of truncated eigenmode analysis motivating new model development). 

This analysis can be extended to further explore stable modes in convection. A significant aspect of RBC concerns the role of boundary conditions (as discussed in many works including Refs.~\cite{Wen20,Wen22,Wang23}). This paper assumed fixed-temperature, no-slip boundary conditions for the entirety of the work. While this is the most common approach for analysis of natural convection, an examination of how the role of stable modes varies with boundary conditions holds potential for rich insights. This work additionally imposed the constraint of 2D dynamics. This serves an appropriate description for many features of RBC, but there are fundamental differences in the evolution of 2D and 3D systems. How stable mode activity varies with the addition of the 3D dimensions also merits exploration. 

Beyond the natural convection of neutral fluids, eigenmode decomposition analysis can be applied to Rayleigh-B\'{e}nard Convection as it arises in magnetized systems that arise in, for example, the Sun's interior. RBC as a model for magneto-convection has an existing precedent in the literature (cf.~\cite{Cresswell23,Aurnou01,Cattaneo03,Cioni00}) and as of yet the eigenmode decomposition analysis has not been applied to such systems. In light of these considerations, this paper serves as an exciting first step into a realm of new investigations in fluid and plasma convection.  

\begin{acknowledgments}
The authors would like to thank A.~E.~Fraser for extensive fruitful discussions throughout this effort. We also would like to thank G.~R.~Donley for her contributions to ideas and preliminary coding efforts in the very early stages of this project. We are also very grateful to the developers of the Dedalus code for their excellent and intuitive programming framework and the regular user support available. This work was conducted using computing resources provided by the Hope College Department of Physics. A.G.L. is grateful for funding support providing by the Clare Booth Luce Research Scholars Program through the Office of the Dean of Natural and Applied Sciences as Hope College. Z.R.W. thanks the Hope College Department of Physics for support. 
\end{acknowledgments}

\bibliography{apssamp}

\end{document}